\documentclass[sn-mathphys-num, iicol]{sn-jnl}

\usepackage{graphics}
\usepackage{graphicx}%
\usepackage{multirow}%
\usepackage{amsmath,amssymb,amsfonts}%
\usepackage{amsthm}%
\usepackage{mathrsfs}%
\usepackage[title]{appendix}%
\usepackage{xcolor}%
\usepackage{textcomp}%
\usepackage{manyfoot}%
\usepackage{booktabs}%
\usepackage{algorithm}%
\usepackage{algorithmicx}%
\usepackage{algpseudocode}%
\usepackage{listings}%

\usepackage{xspace}
\usepackage{dsfont}
\usepackage{float}

\usepackage{stackengine}

\begin{document}

\newcommand{\ELU}{\operatorname{ELU}}
\newcommand{\PyTorch}{\texttt{PyTorch}\xspace}
\newcommand{\Adam}{\texttt{Adam}\xspace}
\newcommand{\AdamMCMC}{\texttt{AdamMCMC}\xspace}
\newcommand{\AdamW}{\texttt{AdamW}\xspace}

\title[Classifier Surrogates]{Classifier Surrogates: Sharing AI-based Searches with the World}


\author*[1]{\fnm{Sebastian} \sur{Bieringer}}\email{sebastian.guido.bieringer@uni-hamburg.de}

\author[1]{\fnm{Gregor} \sur{Kasieczka} \nomail}

\author[2]{\fnm{Jan} \sur{Kieseler} \nomail}

\author[3]{\fnm{Mathias} \sur{Trabs} \nomail}

\affil*[1]{\orgdiv{Institut f\"ur Experimentalphysik}, \orgname{Universit\"at Hamburg}, \orgaddress{\street{Luruper Chaussee 149}, \city{Hamburg}, \postcode{22761}, \country{Germany}}}

\affil[2]{\orgdiv{Institut f\"ur Experimentelle Teilchenphysik}, \orgname{Karlsruher Institut f\"ur Technologie}, \orgaddress{\street{Wolfgang-Gaede-Str. 1}, \city{Karlsruhe}, \postcode{76131}, \country{Germany}}}

\affil[3]{\orgdiv{Institut f\"ur Stochastik}, \orgname{Karlsruher Institut f\"ur Technologie}, \orgaddress{\street{Englerstr. 2}, \city{Karlsruhe}, \postcode{76131}, \country{Germany}}}


\abstract{In recent years, neural network-based classification has been used to improve data analysis at collider experiments.
While this strategy proves to be hugely successful, the underlying models are not commonly shared with the public and rely on experiment-internal data as well as full detector simulations.
We show a concrete implementation of a newly proposed strategy, so-called Classifier Surrogates, to be trained inside the experiments, that only utilise publicly accessible features and truth information. These surrogates approximate the original classifier distribution, and can be shared with the public.
Subsequently, such a model can be evaluated by sampling the classification output from high-level information without requiring a sophisticated detector simulation.
Technically, we show that Continuous Normalizing Flows are a suitable generative architecture that can be efficiently trained to sample classification results using Conditional Flow Matching. 
We further demonstrate that these models can be easily extended by Bayesian uncertainties to indicate their degree of validity when confronted with unknown inputs by the user.
For a concrete example of tagging jets from hadronically decaying top quarks, we demonstrate the application of flows in combination with uncertainty estimation through either inference of a mean-field Gaussian weight posterior, or Monte Carlo sampling network weights.}




\maketitle
\section{Introduction}
\label{intro}

Current experimental work in particle physics, for example by the ATLAS and CMS collaborations, uses deep learning-based taggers to great success~\cite{guest2018_MLatLHC_review, Albertsson_2018_MLinHEP_whitepaper, Radovic2018_MLatHighE_review, Karagiorgi:2021MLinHEP_review}. 
Such models often define unique and essential quantities in the analysis chain, which are hard to understand in terms of physical quantities.
While the performance benefit is apparent, best practices for sharing the analysis as for traditional cut-based analyses~\cite{Kraml2012_recommendations, abdallah2020reinterpretation} are not yet established.
This especially hinders the re-interpretation of experimental results.
Recently, a first set of proposals on sharing neural network-based results has been published~\cite{Araz:2023leshouche_community}.
On the purely technical side, solutions exist for sharing serialized networks~\cite{daniel_hay_guest_2022_6467676_lwtnn,onnx} and 
some first searches shared with serialized models have been made public~\cite{ATLAS:2021fbt, ATLAS:2022ihe, ATLAS:2022zhj, ATLAS:2023azi}.

However, when the model inputs contain features which are not available outside the collaborations or can only be simulated at high computational cost within the collaboration, the benefit of sharing the network weights is limited as results still can either not be reproduced at all, or are very expensive.
Costly and unavailable input features include detector level quantities, such as hits, or highly detector dependent quantities, such as soft jet-substructure variables.
For example, both $b$-taggers of ATLAS and CMS use detector dependent information~\cite{_2016_detector_level, Sirunyan_2018_detector_level} and current research shows the best classification performance is achieved when using detector-level data, rather than only high-level observables~\cite{Karagiorgi:2021MLinHEP_review, Qu:2022mxjParT}.
For these cases, sharing a surrogate model trained to reproduce the classification results from truth-, parton- or reconstruction-level inputs has recently been proposed in discussions at the LHC Reinterpretation Forum and the 2023 PhysTeV workshop at Les Houches~\cite{Araz:2023leshouche_community}. 
We will follow the newly introduced terminology and refer to such models as \textit{Classifier Surrogates}.
In this work
\begin{itemize}
    \item we demonstrate for a concrete example how such a Classifier Surrogate could be constructed and evaluated
    \item and present a novel combination of Continuous Normalizing Flows with Monte Carlo-based Bayesian Neural Networks (BNN) for this purpose.
\end{itemize}

Complementary to sharing the full likelihood or the full statistical model \[p(\text{data} \mid\mu)\text{,} \] a Classifier Surrogate can be used to model dependencies on parameters $\tilde \mu$ that were not explicitly included in the statistical model at the time of the release and are hard to model with public fast simulation tools like Delphes~\cite{de_Favereau_2014}. 
Altering the parameters requires that the released model includes intermediate information, for example distributions of observables that are used in a template fit. 
These might stem from the output $x$ of a complex neural network classifier.
For such distributions, the application of a Classification Surrogate can be beneficial.

In practice, a Classifier Surrogate \[p(x\mid c)\] can be used to predict classifier output from any single-event surrogate input \[c \sim p(c\mid \tilde \mu)\text{.}\] 
This simulation of truth-, parton- or reconstruction-level data allows an arbitrary choice of parameters $\tilde \mu$.
If the simulated event is out-of-distribution (OOD) of the training data of the classifier, the surrogate will predict large uncertainties and thus prevent the practitioner from interpreting the analysis where the classification can not be applied reliably.
For simulated events within the classifiers input range, the surrogate predicts samples from the distribution of viable classifier output.
This output prediction in turn can be used to estimate expectation values in histogram bins of derived observables in full analogy to the processing of the classification results from observed data.
A statistical model for the new parameters \[p(\text{data}\mid \tilde\mu)\] can again be derived from the processed and possibly histogrammed surrogate output, for example by assuming Poisson-distributed bin values.
The surrogate strategy therefore is a truly ``open-world'' approach to sharing a classifier-aided analysis.

The uncertainties from the statistical limitation of the dataset, as well as the the smearing introduced by the detector simulation and reduced information of the input $c$ may also be absorbed into an additional nuisance parameter of the new statistical model.

Depending on the nuisance handling strategy used for classifier training~\cite{Dorigo2020nuisance}, the dependence on the nuisance parameters needs to be included in the surrogate 
\begin{equation*}
p(x\mid c) \rightarrow p(x\mid c,\vartheta)
\end{equation*}
for nuisance-parameterized classifiers or in the corresponding input model 
\begin{equation*}
p(c\mid \tilde\mu) \rightarrow p(c\mid \tilde\mu, \vartheta)
\end{equation*}
for nuisance-invariant approaches.
\newpage

If trained on truth- or parton-level, generating surrogate input events $c \sim p(c|\tilde \mu)$ does not require detector simulation and can thus significantly improve the computational cost of any re-interpretation.
Furthermore, eliminating the detector simulation also removes a major bottle-neck for sharing the results with colleagues, that do not have access to collaboration internal simulation-settings.

We introduce the strategy on the concrete example of a classifier derived from the Particle Transformer~\cite{Qu:2022mxjParT}.
This setup is introduced in Section~\ref{sec:toy}.
In Section~\ref{sec:dist}, we then discuss why a Classifier Surrogate needs to employ a generative architecture and introduce a possible architecture in Section~\ref{sec:arch}.
To model increased uncertainty for unknown inputs, we develop two BNN implementations of the architecture in Section~\ref{sec:BNN}. 
In Section~\ref{sec:res}, we discuss the performance of the surrogate both for data within the distribution of the training data, as well as for data new to the model.
We evaluate calibration and scaling to the tails of the distribution, as well as OOD indication.

\section{Particle Transformer and JetClass Dataset} \label{sec:toy}

As internal taggers of the big collaborations are not available for public study, we choose to emulate the state-of-the-art jet tagger, the Particle Transformer (ParT)~\cite{Qu:2022mxjParT}.
ParT is an attention-based model trained to distinguish $10$ different types of jets using per-particle information and trained on the $100$M JetClass dataset~\cite{qu_2022_6619768_jetclass}.
The features include kinematics, particle identification, and trajectory displacement of every particle in the jet.

From the large initial JetClass dataset as stand-in for the internal collaboration datasets, we distill our toy dataset by calculating transverse momentum, pesudorapidity, scattering angle, jet energy, number of particles, soft drop mass~\cite{Larkoski2014_softdrop} and N-subjettiness~\cite{Thaler2011_subjettiness} for $N=1,...,4$, as well as the output of the full ParT for the regarding event. For the first studies we will restrain the experiments to the first five jet-observables as well as the true top or QCD label as surrogate input.

While learning a surrogate of a multiclassifier is possible by using a generative architecture with a multidimensional output space, we restrict the setup to finding a surrogate for binary classification of top jets.
The toy train and validation datasets contain $1$M jet events each from $Z$-events and hadronic decay of $t\bar{t}$.
To reduce the $10$-dimensional ParT output to a binary classification result, we rescale
\begin{equation*}
    p_\mathrm{t \rightarrow b q q' } 
    = \frac{p_\mathrm{t \rightarrow b q q' }^\mathrm{ParT}}{p_{\mathrm{t \rightarrow b q q' }}^\mathrm{ParT} + p_{\mathrm{Z \rightarrow q q' }}^\mathrm{ParT}}.
\end{equation*}

\section{Detector Smearing Distribution} \label{sec:dist}

\begin{figure}[b]
    \centering
    \includegraphics[width=1\linewidth]{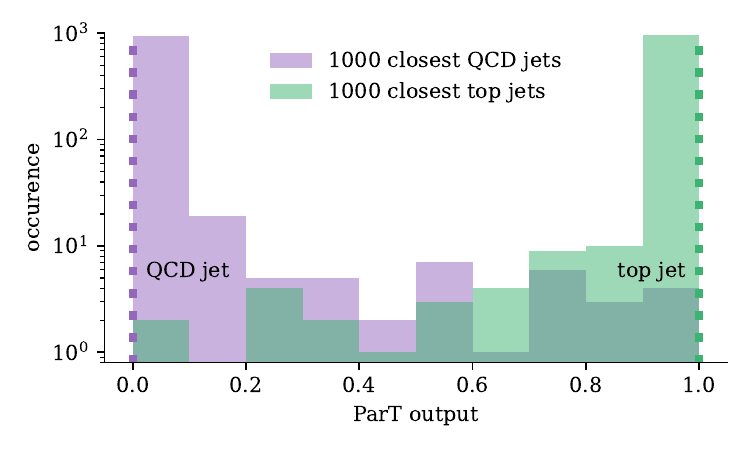}
    \caption{Histograms of the ParT classification results for the $1000$ jet events of the training data, closest to the two arbitrary jets indicated by the dotted lines. 
    Although being classified with varying confidence from detector-level data, the high-level observables $p_T$, $E_\mathrm{jet}$ and $n_\mathrm{const}$ appear identical.}
    \label{fig:detector_smearing}
\end{figure}

Due to the stochasticity of the detector simulation, jets with the same high-level observables can differ a lot on detector-level.
Similarly, jets simulated for identical truth- or parton-level events, would vary significantly.
These jets will thus result in different ParT outputs defining the likelihood per set of high-level observables
\begin{equation}\label{eq:DSD}
    p(\, \underbrace{\addstackgap[3pt]{$\mathrm{ParT}$}}_x \mid \underbrace{p_T, \eta, \phi, E_\mathrm{jet}, n_\mathrm{const}, ...}_c \, ).
\end{equation}
Based on its physical origin, we will also refer to this distribution as the Detector Smearing Distribution.

We can generate a first approximation of this distribution by generating a histogram of the ParT output for the closest points in $p_T$, $E_\mathrm{jet}$ and $n_\mathrm{const}$.
In Figure \ref{fig:detector_smearing} we show this histogram for the $1000$ nearest jet events in the training sample for two arbitrary jet events in the bulk of the transverse momentum distribution at $p_t \approx 530 \, \mathrm{GeV}$.
The imperfect ParT classification introduces an output distribution with tails for events indistinguishable from the high-level features.
Employing a generative architecture as introduced in Section~\ref{sec:arch}, allows us to infer this distribution from the high-level observables.

For the toy setup, we assume the classifier to be constructed invariant for the relevant nuisance parameters~\cite{Dorigo2020nuisance}.
Whenever a nuisance-parameterized classifier is applied, the nuisance parameters need to be included into the likelihood as well.

\section{Neural Density Estimation}\label{sec:arch}
While all flavours of generative models have found numerous applications in high-energy physics, for example in~\cite{Butter22_MLgeneration} and~\cite{Hashemi:2023rgo_generative}, Normalizing Flows can easily and efficiently be applied to infer complex, low-dimensional conditional distributions~\cite{NF_cond_winkler2019learning, NF_cond_radev2020bayesflow}.
For an early application to particle physics, see for example MadMiner~\cite{Brehmer2019MadMiner} and Bayesflow~\cite{NF_cond_radev2020bayesflow,Bieringer:2020CINN_QCD}.
In our tests, coupling block-based Normalizing Flows exhibit great performance for dense phase space regions, but larger deviations when modelling tails of distributions.
To boost the performance of the model we employ Continuous Normalizing Flows (CNF), a generalization of coupling block Flows based on ordinary differential equations (ODE) introduced in Section \ref{ssec:CNF}. 

In Classifier Surrogates, the deficiency of coupling block-based Normalizing Flows to model distribution tails is masked to large extend by the softmax-normalization employed on the classifier, and thus also surrogate, when calculating class probabilities.
We do observe similar performance between both architectures.
However, CNFs are also much more parameter efficient allowing us to reduce the number of parameters needed by a factor of $\approx 20$ at the cost of slower inference time.
As the weights of the surrogate are designed to be shared, and we do expect their use in case studies rather than evaluating on millions of jets, we believe that CNFs are best suited for the application.

\subsection{Continuous Normalizing Flows and Conditional Flow Matching}\label{ssec:CNF}

First introduced in~\cite{CNF_chen2018neural}, CNFs define a transformation $\phi_t: [0,1] \times \mathds{R}^d \rightarrow \mathds{R}^d$ called \textit{flow} dependent on a time variable $t$.
The time variable is the continuous equivalent to the number of a coupling blocks in a coupling block-flow~\cite{flows_rezende2015variational}.
Instead of having multiple flow instances, the dependence of $\phi$ on $t$ is defined through the ODE
\begin{equation}\label{eq:CNF_ODE}
    \frac{\mathrm{d}}{\mathrm{d} t} \phi_t(x) = v_t(\phi_t(x)),\quad \phi_0(x)=x,    
\end{equation}
by the time dependent \textit{vector-field} $v_t\colon [0,1] \times \mathds{R}^d \rightarrow \mathds{R}^d$, which itself is approximated by a deep neural network 
\begin{equation*}
    \tilde v_t(\cdot,\theta)\approx v_t. 
\end{equation*}
While this network can be arbitrarily complex, we stick to fully-connected architectures due to the low dimensionality of the task.
In our case, the flow transforms data from a Gaussian distribution $\mathcal{N}(0,1)$ for $t=0$ into ParT output at $t=1$.
This choice sets the boundaries of the \textit{probability path} $p_t\colon[0,1] \times \mathds{R}^d \rightarrow \mathds{R}_{>0}$ induced by the vector-field trough Equation \eqref{eq:CNF_ODE} and
\begin{equation}\label{eq:COV}
    p_t(x) = p_0\left( \phi_t^{-1}(x)\right) \det \left( \frac{\partial \phi_t^{-1}(x)}{\partial x}\right).
\end{equation}

A standard CNF is trained by solving the ODE Equation \eqref{eq:CNF_ODE} in reverse and minimizing the negative $\log$-likelihood (NLL) of input data at $t=1$.
The computation of this loss objective is expensive, especially for higher dimensional models.

Thus, the authors of~\cite{CFM_lipman2023flow} introduce the Conditional Flow Matching (CFM) objective
\begin{equation}\label{eq:CFM}
    \mathcal{L}_{\mathrm{CFM}}(\theta)=\mathds{E}_{t, q(x_1), p_t(x|x_1)}\left\|u_t(x|x_1)- \tilde{v}_t(x; \theta))\right\|^2
\end{equation}
It reduces the calculation of the optimization criterion to the calculation of a mean-squared error between the network output $\tilde{v}_t(x; \theta)$ and an analytical solution $u_t$ for sampled $t \sim \mathcal{U}(0,1)$, $x_1 \sim q$ and $x \sim p_t(\cdot|x_1)$.
Here, $q$ is the probability distribution of the input data.
A good choice of $u_t$ and corresponding $p_t$ is a Gaussian conditional probability path with mean and variance changing linear in time (optimal transport)~\cite{CFM_lipman2023flow}.
The CFM-loss \eqref{eq:CFM} then reduces even further
\begin{equation}\label{eq:CFM_OT}
\begin{aligned}
    \mathcal{L}_{\mathrm{CFM}}(\theta)
    = \mathds{E}_{t, q(x_1), p(x_0)}
    \Big\| & \left( x_1-\left(1-\sigma_\mathrm{min}\right)x_0\right)  \\
    & - \tilde{v}_t(\sigma_t x_0+\mu_t; \theta))\Big\|^2,
\end{aligned}
\end{equation}
where $\mu_t = tx_1$, $\sigma_t = 1-(1-\sigma_\mathrm{min})t$, $p(x_0) = \mathcal{N}(0,1)$ and $\sigma_\mathrm{min}$ a small parameter, that can be chosen to match the noise level of the training data.

\subsection{Conditional Density Estimation}\label{ssec:conds}

Following the coupling-block Flow based example of~\cite{NF_cond_radev2020bayesflow}, we can extend CNFs to approximate a conditional density 
\begin{equation}\label{eq:COV_cond}
    p_t(x\mid c) = p_0\left( \phi_t^{-1}(x,c)\mid c\right) \det \left( \frac{\partial \phi_t^{-1}(x,c)}{\partial x}\right),
\end{equation}
where the noise distribution is independent of the condition $p_0(\cdot \mid c) = p_0(\cdot)$, by appending the vector of conditions to every layer of the vector field model $\tilde{v}_t(x, c\,; \theta)$. 
For our surrogate, $x$ will be the ParT output and $c$ will be the vector of jet-observables.

\section{Bayesian Neural Networks} \label{sec:BNN}

To indicate the application of the surrogate on data not included in tagger and thus surrogate training, we employ Bayesian Deep Learning.
Through modeling of (or sampling from) a posterior weight distribution \[\pi\left(\theta \mid \mathcal{D}\right),\] these methods give a large spread of predictions for data not included in the loss objective during training.
This posterior distribution is the distribution of weights $\theta$ of the network $\tilde{v}_t(\cdot, \theta)$ given the training data \[\mathcal{D}= \left\{ (x^{(1)},c^{(1)}),(x^{(2)},c^{(2)}),...\right\}.\]
Multiple instances from the weight posterior will form an ensemble of networks with differing weights.
With both being conditional probability distributions, the weight posterior has to be distinguished from the likelihood of classifier output~\eqref{eq:DSD} that is to be inferred by every CNF in the ensemble.
Sections \ref{ssec:VIB} and \ref{ssec:AdamMCMC} introduce two different approaches to connect both distributions.

\subsection{Mean-Field Gaussian Variational Inference (VIB)}\label{ssec:VIB}

A first way to relate the the weigth posterior $\pi\left(\theta \mid \mathcal{D}\right)$ to a CNF is to  approximate it with  an uncorrelated  Normal distribution $\tilde{\pi}(\theta)$~\cite{bbb_blundell2015weight}.
This approximation is usually inferred during optimization of the network, by minimizing the Kullback-Leibler divergence ($D_\mathrm{KL}$) between the posterior and its approximation
\begin{equation}\label{eq:bbb}  
\begin{aligned}
    \mathcal{L}_{\mathrm{VIB}}
    &=D_\mathrm{KL}\left[\, \tilde{\pi}(\theta), \pi\left(\theta  \mid  \mathcal{D}\right)\right] \\
    &=-\int \mathrm{d} \theta \, \tilde{\pi}(\theta) \, \log \pi\left(\mathcal{D} \mid  \theta\right) \\
    & \quad \quad \quad + D_\mathrm{KL}\left[\, \tilde{\pi}(\theta), \pi(\theta)\right]+\text{ constant},
\end{aligned}
\end{equation}
where $\pi(\theta)$ is the prior imposed on the network weights.
Following the construction in~\cite{CFM_butter2023jet}, we bridge the gap between the CFM-loss \eqref{eq:CFM_OT} and the $\log$-likelihood of the data in \eqref{eq:bbb} by introducing a factor $k$ that can be optimized to account for the difference
\begin{equation}\label{eq:bbb_CFM}
    \mathcal{L}_{\mathrm{VIB-CFM}} = \mathds{E}_{\tilde{\pi}(\theta)}\mathcal{L}_{\mathrm{CFM}} 
    + k D_\mathrm{KL} \left[\, \tilde{\pi}(\theta), \pi(\theta)\right].
\end{equation}

\subsection{\AdamMCMC}\label{ssec:AdamMCMC}
While the derivation of the loss \eqref{eq:bbb_CFM} lacks theoretic backing and its optimization can take considerably longer than that of the CFM-loss~\eqref{eq:CFM_OT} alone, the low dimensionality of the Classifier Surrogate problem allows us to directly sample the weight posterior distribution through Markov Chain Monte Carlo (MCMC).

Full Hamiltonian Monte Carlo (HMC) is still often considered the gold-standard for inferring weight posteriors~\cite{izmailov2021bayesian_HMC}.
The large size of the training data however forces us to use stochastic MCMC algorithms. 
As one instance of this class, we choose \AdamMCMC~\cite{bieringer2023adammcmc} due to its easy implementation. 
Competing algorithms, such as stochastic gradient HMC~\cite{chen2014sgHMC} or symmetric splitting HMC~\cite{cobb2021ssHMC}, will likely produce similar results. 

We initialize the \AdamMCMC-chain with a network optimized using the CFM-loss objective \eqref{eq:CFM_OT} and solve the ODE \eqref{eq:CNF_ODE} to determine the negative $\log$-likelihood $\mathcal{L}_\mathrm{NLL}$ of the data for every step of the MCMC from there on.

To ensure detailed balance we employ a Metropolis-Hastings (MH) correction with acceptance rate
\begin{equation}
    \alpha = \frac{\exp \left(-\lambda \mathcal{L}_{\mathrm{NLL}}(\tau_i)\right) q(\theta_i \mid  \tau_i)}{\exp \left(-\lambda \mathcal{L}_{\mathrm{NLL}}(\theta_i)\right) q(\tau_i \mid  \theta_i)}
\end{equation}
for all steps of the chain. 
Through the proportionality  $\pi\left(\theta \mid \mathcal{D}\right) \propto -\mathcal{L}_{\mathrm{NLL}}$ (Bayes formula) the acceptance step guarantees sampling from the weight posterior.
Here, the parameter $\lambda$ gives the inverse temperature of the tempered posterior distribution sampled from.
The proposed weights $\tau_i$ are drawn from a proposal distribution centered on a gradient descent step 
\begin{equation}
    \tilde\theta_{i+1} = \mathrm{Adam}(\theta_i,  \mathcal{L}_{\mathrm{NLL}}(\theta_i))
\end{equation}
calculated using the \Adam algorithm~\cite{adam_Kingma2014AdamAM}. 
We can use the momentum terms of the update to ensure high acceptance rates by smearing the proposal distribution in the direction of the last update
\begin{equation}
\begin{aligned}
    \tau_i \sim \, & q(\cdot\mid \theta_i) \\
    & = \mathcal{N}(\tilde\theta_{i+1},\sigma^2\mathrm{1}+(\tilde\theta_{i+1} - \theta_i)(\tilde\theta_{i+1} - \theta_i)^\top).
\end{aligned}
\end{equation}

To efficiently run this algorithm, we evaluate the NLL on batches of data. 
For proofs on convergence and invariant distribution of this algorithm, we refer to~\cite{bieringer2023adammcmc}.

\section{Results} \label{sec:res}

To learn the Detector Smearing Distribution from data, we found a CNF with only $3$ multi-layer perceptrons (MLPs) with $3$ layers of dimension $64$ and $\ELU$ activation to be sufficient.
The condition and time variable $t$ are concatenated to every MLP input, totalling in $31617$ network parameters.
Converting to VIB as in~\cite{bbb_blundell2015weight}, doubles the number of parameters.
We train on a balanced set of $4$M jets in batches of $131072$ for $4000$ epochs using \Adam~\cite{adam_Kingma2014AdamAM} with a constant learning rate of $10^{-3}$.
As loss objective, we use the CFM-loss as introduced in Equations \eqref{eq:CFM_OT} and \eqref{eq:bbb_CFM} respectively.
To achieve good coverage, we choose $c=100$ and $\lambda = 50$ from a course grid search over multiple orders of magnitude.

We run the \AdamMCMC chain for another $1000$ epochs with the learning rate reduced to $5\cdot10^{-6}$ and $\sigma=0.05$. 
For the sampled posterior we always report the results from CFM-optimization in solid lines and the uncertainty calculated as the $\min$-$\max$-envelope of $10$ drawings and for the learned approximation (VIB) we give the mean and the $\min$-$\max$-envelope over $11$ sets of weights.

Using a fully-connected architecture, the sampled networks, either from the VIB-approximation or MCMC, can be easily exported as a serialized file using \textsc{ONNX}~\cite{onnx} at only 0.3 MB per instance. 
The the ODE defined by the network remains to be solved at inference time.

\subsection{In-Distribution}

\begin{figure}[b]
    \centering
    \includegraphics[width=1\linewidth]{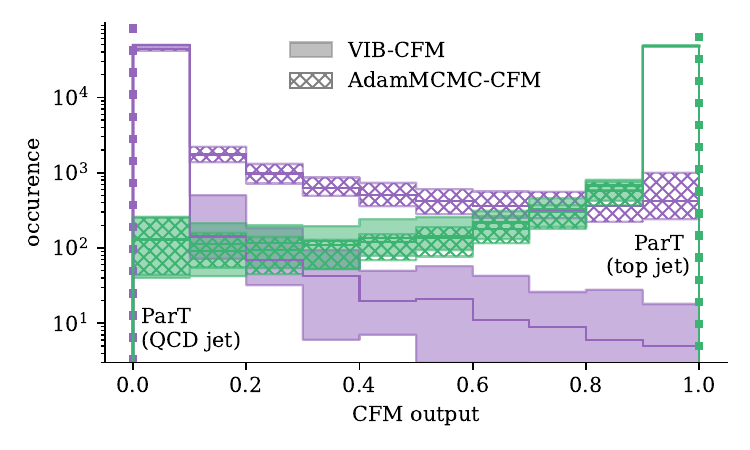}
    \caption{Histograms of $50000$ samples drawn form the Detector Smearing Distributions learned with a CFM-model.
    Uncertainties are generated by drawing the samples from $11$ points sampled from the network posterior approximation or chain.
    The ParT-output for the arbitrary QCD and top jet used as condition is indicated with dotted lines.
    Both jet events are the same as for Figure \ref{fig:detector_smearing}.
    }
    \label{fig:learned_detector_smearing}
\end{figure}

We can use the trained CNFs to generate another approximation of the Detector Smearing Distribution by performing the forward direction starting at different points in latent space but for the same high-level features.
Figure~\ref{fig:learned_detector_smearing} shows histograms of the generated data for the same arbitrary jet events as Figure~\ref{fig:detector_smearing}.

We can see similar distributions for the approximation with CNFs as for the histograms of the closest events.
The biggest discrepancy occurs between the distribution for the QCD jet obtained using \AdamMCMC and VIB. 
It can be attributed to the difference between the model at initialization of the \AdamMCMC chain and the posterior mean output of VIB.
The initialization can be adapted to accommodate desired behaviours, if well defined, by choosing between different epochs of the CFM-optimization. 
Furthermore, increasing the chain length decreases the dependence of the ensemble output on the initialization overall.

\subsubsection{Uncertainty Calibration}

\begin{figure}
    \centering
    \includegraphics[width=0.8\linewidth]{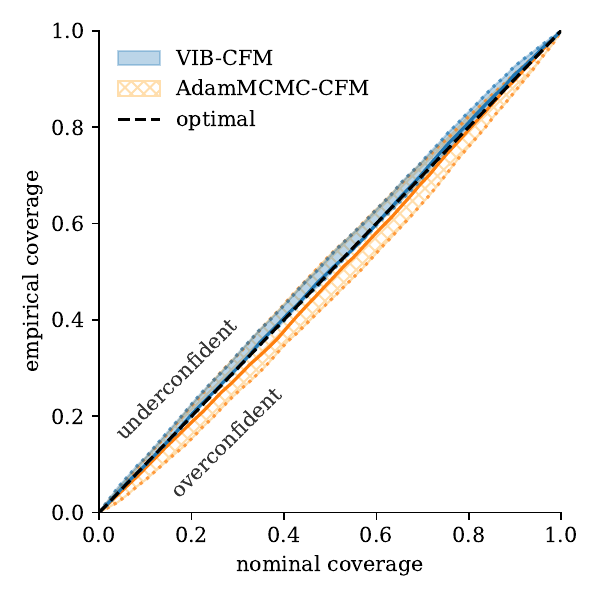}
    \caption{Empirical over nominal coverage calculated by taking $1000$ samples from the learned Detector Smearing Distribution for $10000$ jet events each.
    Uncertainties again are calculated from $11$ points of the network posterior.
    }
    \label{fig:cal}
\end{figure}

To find out whether surrogate predictions using \AdamMCMC are in general conservative, we need to look at the calibration of the estimated Detector Smearing Distributions for multiple events, here $10000$. 
Per event we take $1000$ samples from the inferred distribution and calculate $q$-quantiles for 50 values of $q$ (nominal coverage) linearly spaced between $0$ and $1$.
We then evaluate the empirical coverage, that is the fraction of corresponding ParT output within the respective quantile of the inferred distribution.
The calibration is perfect when nominal and empirical coverage agree.
Figure~\ref{fig:cal} shows very good calibration for both methods, where \AdamMCMC in fact tends to be slightly more confident than VIB approximations.

\subsubsection{Epistemic Uncertainty}
 
\begin{figure*}
    \centering
    \includegraphics[width=\textwidth]{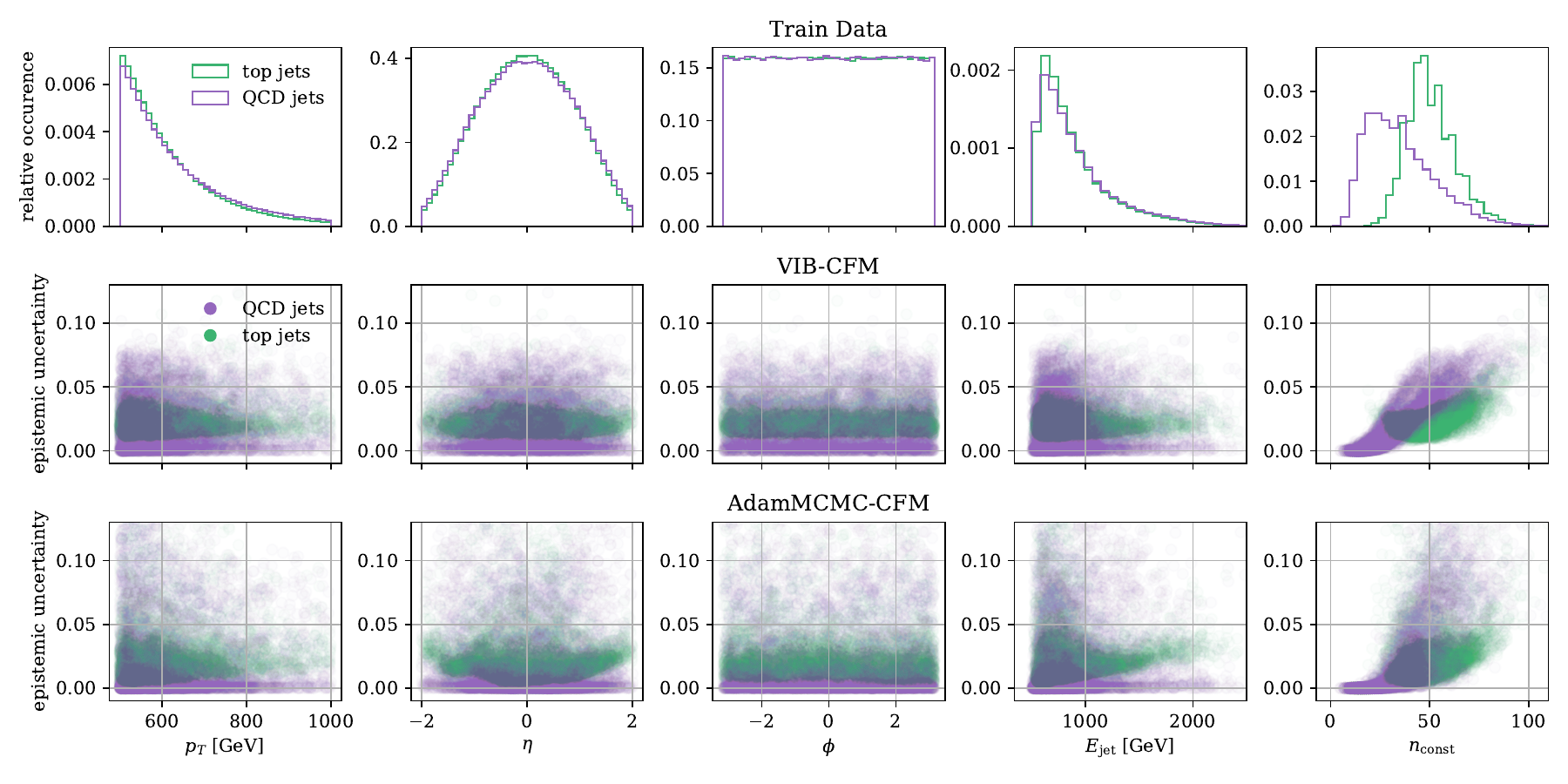}
    \caption{Epistemic uncertainty calculated from the mean difference between 11 points from the network posterior over $1000$ samples drawn from the respective learned Detector Smearing Distribution for each event of the validation set.
    The uncertainty shows a clear scaling towards the edges of the train data, as well es in regions where $n_\mathrm{const}$ is uninformative.
    }
    \label{fig:err}
\end{figure*}

In contrast to uncertainty due to noisy data resulting in the Detector Smearing Distributions, epistemic uncertainty is the uncertainty encoded in the variations within the ensembles of network weights induced by data sparsity. 
For a further dive into the behaviour of the epistemic uncertainty, we calculate the mean distance of the maximum discrepancy between instances of the network posterior
\begin{equation}\label{eq:err}
    \delta_\mathrm{epis} = \frac{1}{n_\mathrm{stat}} \sum_{i=0}^{n_\mathrm{stat}} \max_{p(\theta|\mathcal{D})} \phi_{1,\theta}(x_i) -\min_{p(\theta|\mathcal{D})} \phi_{1,\theta}(x_i)
\end{equation}
for a total of $n_\mathrm{stat}=1000$ points drawn from the Gaussian latent space $x_1,...,x_{n_\mathrm{stat}} \sim \mathcal{N}(0,1)$.
Ideally, this error estimate is large for sparsely populated areas of the high-level feature space and small in the bulk of the distribution.
To investigate this behaviour, we plot a histogram of the high-level features of the training data as well  as $\delta_\mathrm{epis}$ for $10000$ jet events chosen at random from a test set for both methods in Figure~\ref{fig:err}.

The most instructive panels show the dependence of the error estimate on the number of constituents in the jet $n_\mathrm{const}$, which is the most descriptive input feature.
We can see high uncertainties occurring in the regions where the distributions for QCD and top jets overlap in the training data.
These are events that can not easily be attributed to one of the two classes by the five high-level observables alone, resulting in high uncertainties.
These events also make up the high-error bulk when plotted over the other high-level features. 

For every tailed distribution, we can also see an increase of the error estimate for top jet predictions towards the edges of the data.
This behaviour is stronger for \AdamMCMC than for VIB at the cost of higher uncertainties overall.

The same behaviour is not observed for QCD jets.
This again can be traced back to the distribution of $n_\mathrm{const}$.
The distribution of the number of particles of top jets is fully within the support of the one for QCD jets inducing high epistemic uncertainties for both highly and lowly populated jets.
On the other hand, the distribution of top jets does not include events with as few particles as for QCD jets, allowing a perfect classification of these jets that dominates the low uncertainty edge of the plotted cloud.

\subsubsection{Adding Informative Features}

Another measure for the informative value of a Detector Smearing Distribution generated by a Classifier Surrogate is the predicted accuracy

\begin{equation}\label{eq:acc}
     \hat{a}
    = \frac{1}{n_\mathrm{stat}} \sum_{i=0}^{n_\mathrm{stat}} 
    \left\{
\begin{array}{ll}
\mathbf{1}_{[0.5,1]}\left(\phi_{1,\theta}(x_i)\right) \text{ for top jets} \\
\mathbf{1}_{[0,0.5)}\left(\phi_{1,\theta}(x_i)\right) \text{ for QCD jets} \\
\end{array}
\right.\\
\end{equation}
per jet event, with $\mathbf{1}_A(x)$ the indicator function of set $A$.
The cut value of $0.5$ is arbitrary and can be chosen in line with the experimental analysis.
Our choice reflects the requirement to yield symmetric output distributions in case of uninformative high-level input.

\begin{figure}[b]
    \centering
    \includegraphics[width=\linewidth]{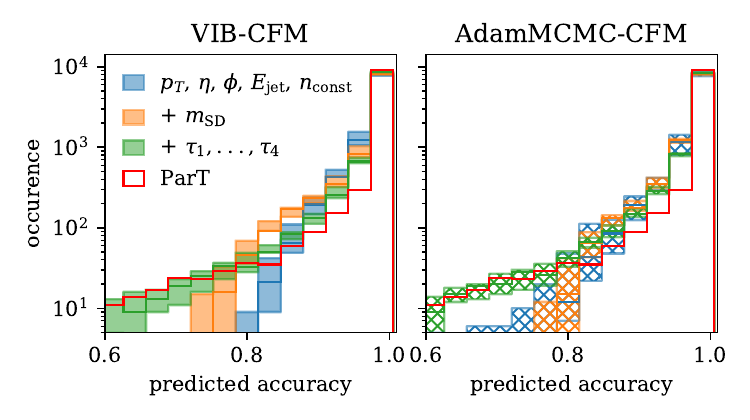}
    \caption{Accuracy of $1000$ ParT outputs predicted for each of $10000$ jet events.
    The different colors indicate the output of CNFs conditioned with increasing amount of features and thus provided with more information during inference.
    A histogram of the probabilistic ParT prediction itself is given in red.
    }
    \label{fig:acc_more_info}
\end{figure}

Figure~\ref{fig:acc_more_info} shows histograms of the predicted accuracy for $10000$ jet events chosen at random from the full balanced test set.
The distributions are generated from surrogates using the five high-level jet features from before, as well as for surrogates including the soft drop mass $m_\mathrm{SD}$ and the $N$-subjettiness for $N\in \{1,..,4\}$.
Naively, we assume that adding more information will lead to more certain predictions and thus will shift the distributions towards high accuracy values.
In the highest value bin, the information hierarchy is well reproduced, with the highest number of input features leading to the highest number of certain outputs.
In the range of $0.85$ to $1$, more informative input leads to fewer predictions in line with the naive assumption.
For less certain predictions, a different effect can be observed.
Increasing the information in the conditions allows the network to better model the ParT output, which features long tails of individual false positives and events predicted with low confidence.
Thus, the Jensen-Shennon divergence between the histograms of surrogate and ParT output (Table \ref{tab:jsd_more_info}) decreases with increasing number of input features.

\begin{figure*}[t]
    \centering
    \includegraphics[width=\textwidth]{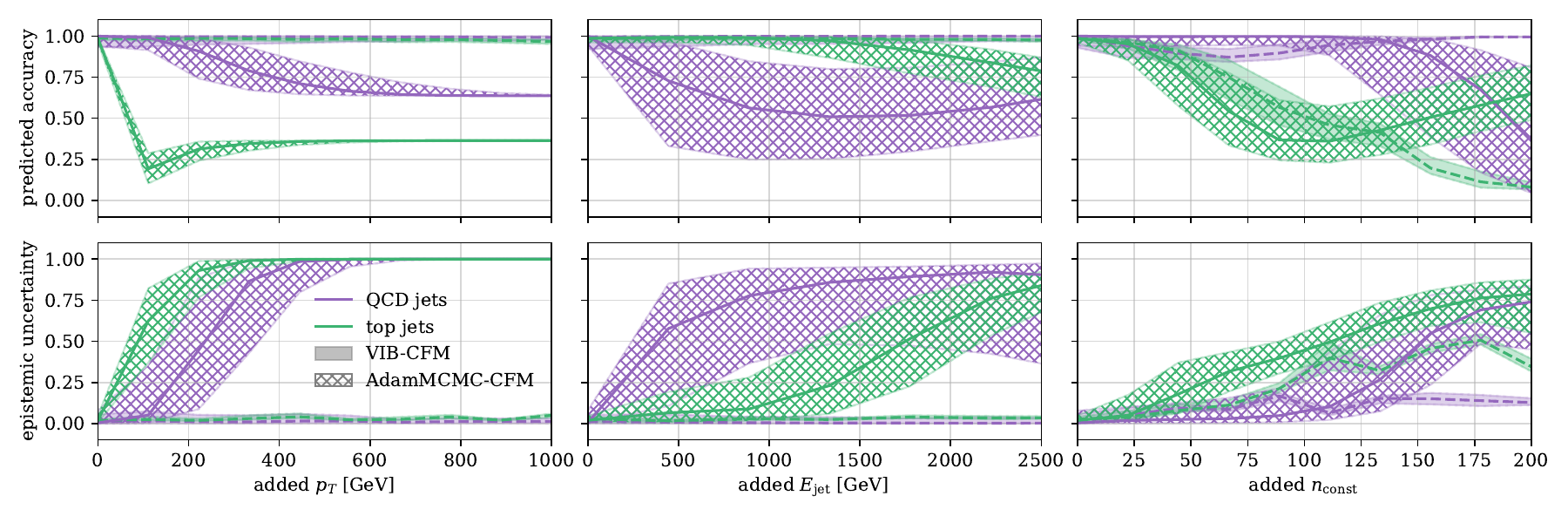}
    \caption{Behaviour of the Classifier Surrogate predictions for distorted input over the size of the distortion artificially added to the given jet-observable for $1000$ random events.
    The first row shows the accuracy predicted per event while the second row gives the mean epistemic uncertainty per event.
    Solid lines give the median of the set of events.
    The shaded and structured areas indicate the $10\%$-$90\%$-quantile envelope of the VIB and \AdamMCMC ensemble respectively.
    }
    \label{fig:odd}
\end{figure*}

\begingroup
\setlength{\tabcolsep}{4pt} 
\renewcommand{\arraystretch}{1} 
\begin{table}[t]
    \centering
    \begin{tabular}{c|cc}
        JSD & VIB-CFM & \AdamMCMC-CFM\\ \midrule
         $p_T$, $\eta$, $\phi$, $E_\mathrm{jet}$, $n_\mathrm{const}$ & $0.174 \pm 0.018$ & $0.147 \pm 0.036$\\
         + $m_\mathrm{SD}$ & $0.134 \pm 0.023$ & $0.160 \pm 0.013$\\
         + $\tau_1,...,\tau_4$ & $0.080 \pm 0.009$ & $0.097 \pm 0.007$\\
    \end{tabular}
    \vspace{0.3cm}
    \caption{Jensen-Shannon-divergence between the histograms of predicted accuracies of Classifier Surrogates with different input features (Figure \ref{fig:acc_more_info}) and the actual accuracy distribution of the ParT.}
    \label{tab:jsd_more_info}
\end{table}
\endgroup

\subsection{Out-of-Distribution}

Although including an epistemic uncertainty into the evaluation this far is a nice feature to gauge uncertainties in the tail regions of the data, the true value of BNNs is indicating input that is outside the distribution of the training data by assigning high uncertainties.
We use the introduced measures \eqref{eq:err} and \eqref{eq:acc} to show the behaviour of the BNN surrogates for OOD data generated when artificially increasing the values for one jet-observable.

We produce OOD data by selecting $1000$ jet events from the test set at random and increasing the values of a single jet-feature by adding a constant value.
We perform this distortion for $3$ dimensions,  $p_T$, $E_\mathrm{jet}$ and $n_\mathrm{const}$, and $10$ values each.
Again, we report the accuracy and error estimate calculated from $n_\mathrm{stat}=1000$ points of the learned Detector Smearing Distribution.

The first row of Figure~\ref{fig:odd} shows the mean accuracy predicted for the OOD data by the different drawings from the weight posterior.
The envelope and solid line give the $10\%$- and $90\%$-quantile and the median over the set of events.
When adding an unphysical offset to the features, we can see the mean predicted accuracy of the \AdamMCMC ensemble rapidly drops.
Optimally, the network predicts $0.5$ when all inputs are outside the training interval to indicate equal confidence of both classes.
The ensemble seems to be able to detect most outliers, but  only indicates large distortions of $E_\mathrm{jet}$ for top jets and of $n_\mathrm{const}$ for QCD jets.

The  predicted accuracy of the VIB samples does not exhibit any dependence on the increasing offset in the OOD data. 
It is sensitive only to the number of jet constituents for top jets.

In the second row, we show the error estimate based on the difference between highest and lowest proposed output in the ensemble, see Equation~\ref{eq:err}.
This measure captures the differences in the output and thus the encoded uncertainty directly.
We expect increasing uncertainties for increasing offset.
Only the \AdamMCMC ensemble shows this behaviour, for all three disturbed input dimensions, while VIB once again is only sensitive to OOD inputs in the particle number.
While the predicted accuracy did not capture the decreasing confidence for distorted $E_\mathrm{jet}$ of top jets well, the error estimate clearly indicates the unknown inputs.
Similarly, distortions in $n_\mathrm{const}$ of QCD jets appear earlier in this measure.

\section{Conclusion}

In this paper, we proposed a first architecture for training Classifier Surrogates, which are models describing the output of a deep neural network classification based on detector-level information from high-level jet-observables and truth information.
We show that the resulting Classifier Surrogates are well calibrated and scale with the amount of information provided.
A combination with Monte Carlo generated samples from the networks Bayesian weight posterior allows for stable uncertainty quantification, that incorporates the density of the training data towards the edges.
The predicted uncertainty reliably indicates unknown inputs.

This approach should next be implemented by the large experimental collaborations to allow the statistical re-interpretation of analysis results.

\backmatter

\bmhead{Acknowledgements}

We thank the organisers and participants of the Les Houches -- PhysTeV 2023 workshop for the stimulating environment which led to the idea for this work. 

Furthermore, we thank Joschka Birk for helping with data preparation in the early stages of the work.

\section*{Declarations}

\bmhead{Code}

The training and evaluation code is available from \url{https://github.com/sbieringer/ClassificationSurrogates} and an example on handeling the JetClass dataset is given in \url{https://github.com/joschkabirk/jetclass-top-qcd}. 

\bmhead{Funding}

SB is supported by the Helmholtz Information and Data Science Schools via DASHH (Data Science in Hamburg - HELMHOLTZ Graduate School for the Structure of Matter) with the grant HIDSS-0002.
SB and GK acknowledge support by the Deutsche Forschungsgemeinschaft (DFG) under Germany’s Excellence Strategy – EXC 2121  Quantum Universe – 390833306.
JK is supported by the Alexander-von-Humboldt-Stiftung.
This research was supported in part through the Maxwell computational resources operated at Deutsches Elektronen-Synchrotron DESY, Hamburg, Germany. 

\bibliography{literature}

\end{document}